\pdfoutput=1

\documentclass[11pt,twoside,a4paper,cmspaper,final,collab]{cms-tdr}
\def\svnVersion{344469}\def\cmsCernNoTag{CERN-PH-EP/2015-302}\def\cmsCernDate{\today}\def\cmsMessage{Published in Physics Letters B as \href{http://dx.doi.org/10.1016/j.physletb.2016.05.044}{\doi{10.1016/j.physletb.2016.05.044.}}}

\begin{document}\cmsNoteHeader{HIN-15-002}

\hyphenation{had-ron-i-za-tion}
\hyphenation{cal-or-i-me-ter}
\hyphenation{de-vices}
\RCS$Revision: 344362 $
\RCS$HeadURL: svn+ssh://svn.cern.ch/reps/tdr2/papers/HIN-15-002/trunk/HIN-15-002.tex $
\RCS$Id: HIN-15-002.tex 344362 2016-05-26 07:40:44Z azsigmon $
\newlength\cmsFigWidth
\ifthenelse{\boolean{cms@external}}{\setlength\cmsFigWidth{0.85\columnwidth}}{\setlength\cmsFigWidth{0.4\textwidth}}
\ifthenelse{\boolean{cms@external}}{\providecommand{\cmsLeft}{top\xspace}}{\providecommand{\cmsLeft}{left\xspace}}
\ifthenelse{\boolean{cms@external}}{\providecommand{\cmsRight}{bottom\xspace}}{\providecommand{\cmsRight}{right\xspace}}
\cmsNoteHeader{HIN-15-002}
\title{Study of Z boson production in pPb collisions at $\sqrt{{s_{NN}}}=5.02$\TeV}

\author{The CMS Collaboration}

\date{\today}

\abstract{
The production of Z bosons in pPb collisions at $\sqrt{s_\mathrm{NN}}=5.02$\TeV is studied by the CMS experiment via the electron and muon decay channels.
The inclusive cross section is compared to pp collision predictions, and found to scale with the number of elementary nucleon-nucleon collisions.
The differential cross sections as a function of the Z~boson rapidity and transverse momentum are measured.
Though they are found to be consistent within uncertainty with theoretical predictions both with and without nuclear effects, the forward-backward asymmetry suggests the presence of nuclear effects at large rapidities.
These results provide new data for constraining nuclear parton distribution functions.
}

\hypersetup{%
pdfauthor={CMS Collaboration},%
pdftitle={Study of Z boson production in pPb collisions at sqrt(s[NN]) = 5.02 TeV},%
pdfsubject={CMS},%
pdfkeywords={CMS, physics, heavy ions}}

\maketitle

\section{Introduction}

Electroweak boson production is an important benchmark process in high-energy particle phy\-sics. The production of Z and W~bosons has been extensively studied at hadron and e$^+$e$^-$ colliders, at various collision energies. The latest measurements in pp collisions at the LHC~\cite{Aad:2011dm, Aad:2011gj, CMS:2011aa, Chatrchyan:2011wt, Chatrchyan:2014mua, Khachatryan:2015oaa, Aaij:2012mda, Aaij:2012vn} are well described by the standard model using higher-order perturbative quantum chromodynamics (QCD) and parton distribution functions (PDFs).

With its large center-of-mass energy and high luminosity, the LHC enables for the first time the study of Z and W~boson production in heavy ion collisions. Electroweak bosons are unmodified by the hot and dense medium created in nucleus-nucleus collisions, and their leptonic decays are of particular interest since leptons pass through the medium without being affected by the strong interaction. Both the Z and W~boson production were measured by the ATLAS~\cite{Aad:2012ew, Aad:2014bha} and the CMS~\cite{Chatrchyan:2014csa, Chatrchyan:2012nt} experiments using PbPb collisions taken in 2010 and 2011 at a center-of-mass energy per nucleon pair of $\sqrt{s_\mathrm{NN}}=2.76$\TeV, confirming that the production cross section scales with the number of elementary nucleon-nucleon collisions with a precision of about 10\%.

However, in nuclear collisions, the production of electroweak bosons can be affected by the initial conditions of the collision. The free-proton PDFs are expected to be modified for protons bound in the Pb nucleus, which, together with the fact that the nucleus contains neutrons as well as protons (isospin effect), can modify the observed cross sections as compared to pp collisions. Various groups have studied the nuclear modification of PDFs, and several results are available at next-to-leading-order (NLO) precision in QCD~\cite{deFlorian:2011fp, Eskola:2009uj, Hirai:2007sx}. These results are obtained by global fits to the available deep inelastic scattering and Drell--Yan data, which constrain the nuclear PDFs (nPDFs) in the region of parton longitudinal momentum fraction $x>10^{-2}$ and four-momentum transfer squared $Q^2<(10\GeV)^2$.

The production of electroweak bosons in proton-nucleus collisions at the LHC provides an opportunity to study the nPDFs at the high $Q^2 \approx (100\GeV)^2$ and lower $x$ phase space region~\cite{Salgado:2011pf}. The CMS experiment made the first measurement of W boson production in pPb collisions~\cite{Khachatryan:2015hha}. Deviations from the current expectations for PDFs were observed, showing the need for including W~boson data in nPDF global fits. Furthermore, the dijet pseudorapidity distribution measured in pPb collisions by CMS~\cite{Chatrchyan:2014hqa} and the Z~boson production in pPb collisions measured by ATLAS~\cite{Aad:2015gta} show better agreement with modified PDFs. Deviations from pp expectations were also seen with charged hadrons~\cite{Khachatryan:2015xaa}.

Various models predict different nuclear modifications of the Z~boson production cross section~($\sigma$) as a function of transverse momentum ($\pt$) and rapidity in the nucleon-nucleon center-of-mass frame ($y_\mathrm{cm}$)~\cite{Guzey:2012jp, Kang:2012am, Kartvelishvili:1995fr, Paukkunen:2010qg, Vogt:2000hp}. Processes mediated by a virtual photon and interference effects are also considered as part of the Z~boson signal. The rapidity distribution of Z bosons is particularly sensitive to the parton content of the interacting nucleons. Consequently, the symmetric  rapidity spectrum of the Z bosons in the center-of-mass frame of pp collisions is modified by nuclear effects in pPb collisions~\cite{Paukkunen:2010qg}. This can be quantified through measurements of the forward-backward asymmetry in the center-of-mass frame:
\begin{linenomath}
\begin{equation}\label{eq:fb}
R_\mathrm{FB}(y_\mathrm{cm}) = \frac{\rd\sigma(+y_\mathrm{cm})/\rd y_\mathrm{cm}}{\rd\sigma(-y_{\rm cm})/\rd y_{\rm cm}},
\end{equation}
\end{linenomath}
where by convention positive rapidity values correspond to the direction of the incoming proton.

The aim of this paper is to study the $\cPZ\to\ell\ell$ process (where $\ell$ represents either muons or electrons) and to measure the production cross section as functions of rapidity and transverse momentum. The typical quark momentum fraction probed in the Pb nucleus is given by $x=M_{\Z}/\sqrt{s_\mathrm{NN}}\re^{-y_\mathrm{cm}}$, and thus with $0.002<x<0.3$ in the measured range of $-2.8<y_\mathrm{cm}<2.0$. These measurements will help to constrain the parton content of the nucleons in the nucleus.

\section{Experimental setup, data selection and reconstruction}

A detailed description of the CMS detector and its coordinate system can be found elsewhere~\cite{Chatrchyan:2008zzk}. Its central feature is a superconducting solenoid with internal diameter of 6\unit{m}, providing a magnetic field of 3.8\unit{T}. Within the solenoid volume are a silicon pixel and strip tracker, a lead tungstate crystal electromagnetic calorimeter (ECAL), and a brass and scintillator hadron calorimeter (HCAL). Muons are detected in the pseudorapidity range $\abs{\eta_{\text{lab}}}< 2.4$ using gas-ionization detectors embedded in the steel return yoke outside the solenoid. Electrons are measured in the ECAL that consists of 75\,848 lead tungstate crystals providing a coverage in the barrel region of $\abs{\eta_{\text{lab}}}<1.48$  and in the two endcap regions of $1.48<\abs{\eta_{\text{lab}}}<3.00$. Extensive forward calorimetry complements the coverage provided by these barrel and endcap detectors. CMS has a two-level trigger system. The first level, composed of custom hardware processors, uses information from the calorimeters and muon detectors to select the most interesting events. The high-level trigger processor farm further decreases the event rate before data storage.

The analysis is performed using the pPb collision data taken at the beginning of 2013 and corresponding to an integrated luminosity of $34.6\pm1.2\nbinv$~\cite{CMS:2014uoa}. The beam energies were 4\TeV for protons and 1.58\TeV per nucleon for lead nuclei, resulting in a center-of-mass energy per nucleon pair of $\sqrt{s_\mathrm{NN}}=5.02$\TeV. As a consequence of the energy difference between the colliding beams, the nucleon-nucleon center-of-mass frame is not at rest with respect to the laboratory frame. Massless particles emitted at rapidity $y_\mathrm{cm} = 0$ in the nucleon-nucleon center-of-mass frame will be detected at $y_{\text{lab}} = -0.465$ (clockwise proton beam) or +0.465 (counterclockwise proton beam) in the laboratory frame. The results presented here are expressed in the center-of-mass frame with the proton-going side defining the region of positive $y_\mathrm{cm}$ values, to respect the usual convention of the proton fragmentation region being probed at positive rapidity. The direction of the higher energy proton beam was initially clockwise and was then reversed, producing two comparable datasets.

During data taking, muon and electron triggers were employed to select and record all events with high-\pt leptons. The measurements in the muon final state are based on a sample obtained by requiring at least one muon with $\pt$ greater than 12\GeVc. The muon candidates are reconstructed with an algorithm that combines information from both the silicon tracker and the muon system~\cite{Chatrchyan:2012xi}. Background muons from cosmic rays and heavy-quark semileptonic decays are rejected by applying a set of quality criteria to each muon, based on previous studies of the performance of the muon reconstruction~\cite{Chatrchyan:2012xi}. The muons are selected by requiring at least two muon stations to be matched to the muon track, a low $\chi^2/$ndf of the global fit, a minimum number of tracker layers and pixel hits, and finally, a maximum distance from the primary vertex in the transverse and longitudinal direction.

The electron measurements are based on a candidate photon or electron sample collected by requiring at least one ECAL transverse energy deposit of $\ET>15$\GeV and online identification criteria that are looser than the electron selection applied offline. Electrons are reconstructed by matching ECAL clusters to tracks measured in the silicon tracker. This matching is used to differentiate electrons from photons~\cite{Chatrchyan:2013dga}. The identification criteria are chosen to match those used for pp collisions~\cite{Khachatryan:2015hwa}. The electrons are selected by requiring a match between the $\eta$ and $\phi$ coordinates of the track and the ECAL cluster, a narrow width of the ECAL cluster in $\eta$, a low HCAL energy measured in the ECAL cluster direction and by rejecting electrons with a partner track consistent with a photon conversion. In this measurement, no isolation requirements are imposed on the leptons.

\section{Analysis procedure}

The Z boson production cross section is calculated using the following equation:
\begin{linenomath}
\begin{equation}\label{eq:xsec}
\sigma = \frac{S - B}{\alpha \, \epsilon \, L_\text{int}},
\end{equation}
\end{linenomath}
where $S$ is the number of Z candidates, $B$ is the estimated background, $\alpha$ is the acceptance, $\epsilon$ is the efficiency, including correction factors derived from data, and $L_\text{int}$ is the integrated luminosity. The phase space region considered in the analysis is defined by requiring two leptons with $\pt^{\ell} > 20$\GeVc and with pseudorapidity in the laboratory frame $\abs{\eta_{\text{lab}}^{\ell}} < 2.4$ in order to ensure that the triggers are maximally efficient and are within the geometrical coverage of the muon detectors. This fiducial region of the measurement is extrapolated to the full phase space over $\pt^{\ell}$ and $\eta_{\text{lab}}^{\ell}$ by the acceptance correction. Each component of Eq.~(\ref{eq:xsec}) is presented and systematic uncertainties summarized below.

\subsection{Signal and background}

\begin{figure}[thbp]
 \centering
    \includegraphics[width=0.46\textwidth]{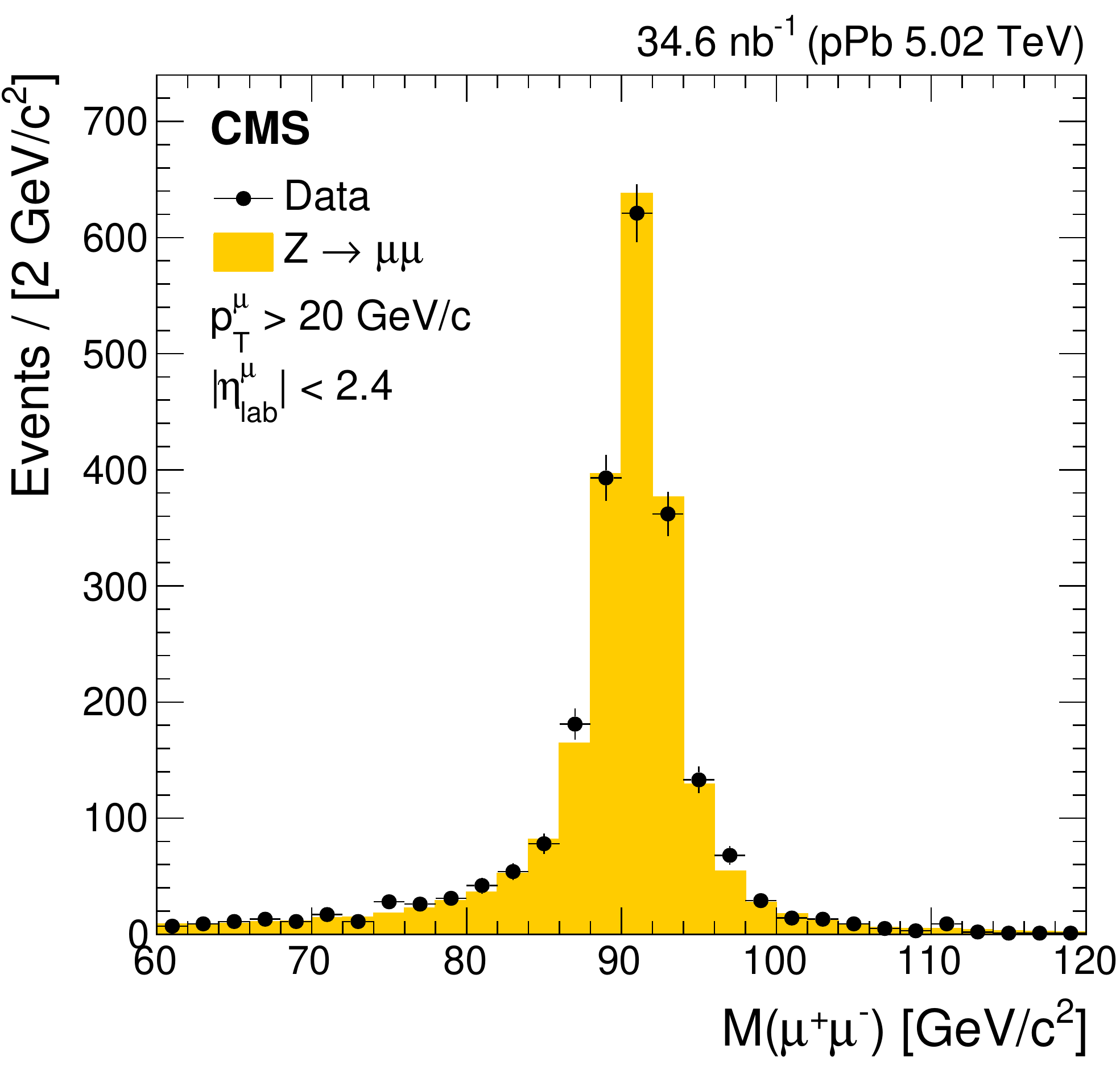}
    \includegraphics[width=0.46\textwidth]{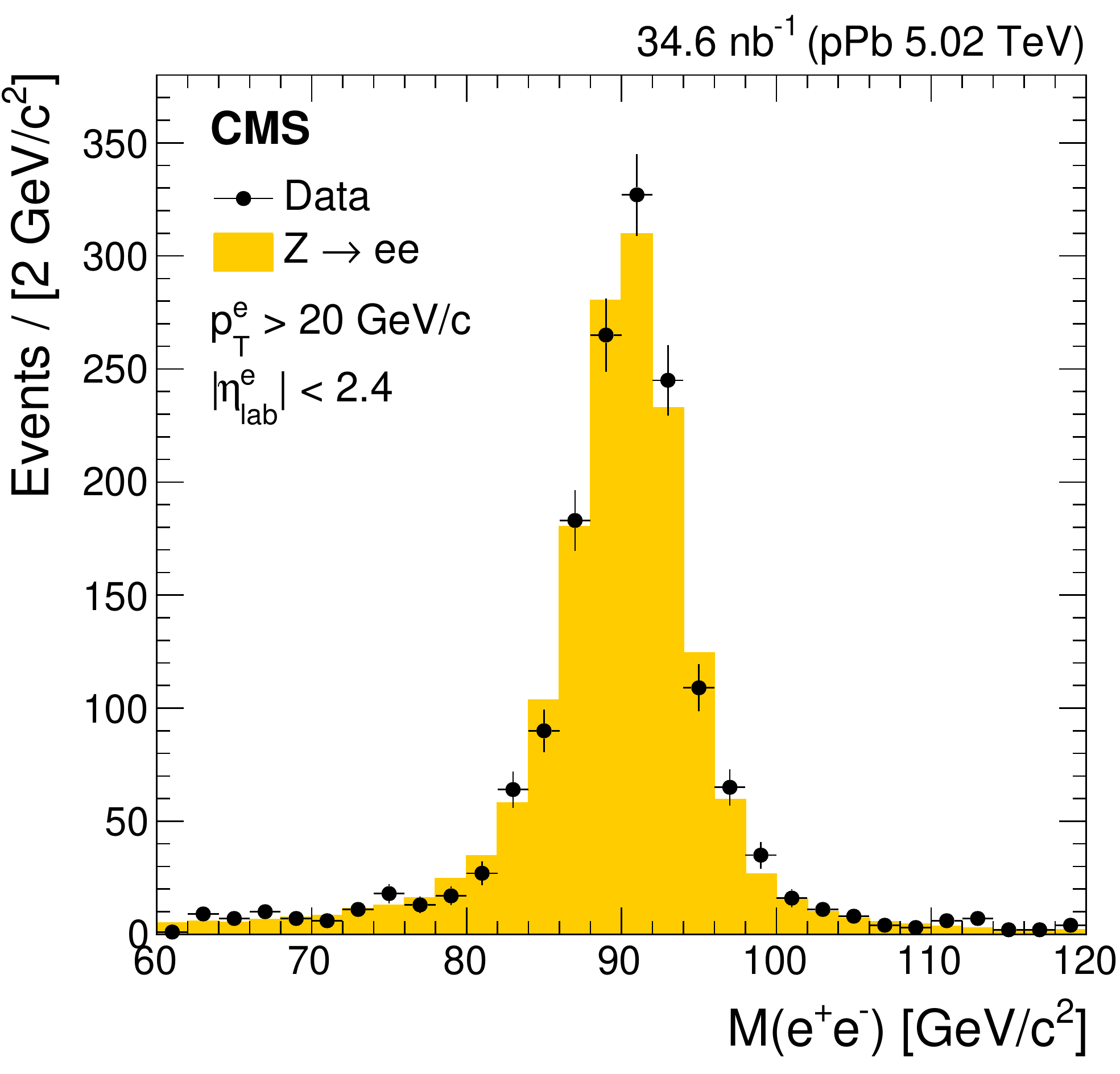}
    \caption{Invariant mass of selected muon (left) and electron (right) pairs compared to \PYTHIA6\textsc{+hijing} simulated $\text{pN}\to\cPZ\to\ell\ell$ events with $\mathrm{N} = (\Pp, \Pn)$ according to the number of nucleons in the Pb nucleus. The MC sample is normalized to the number of events in the data.}
    \label{fig:massplot}
\end{figure}

The Z candidate events are selected by requiring a same-flavor, oppositely-charged lepton pair with an invariant mass in the 60--120\GeVcc range, where both leptons satisfy the acceptance and quality requirements and at least one of them corresponds to the lepton that triggered the event. Figure~\ref{fig:massplot} shows the invariant mass distribution of the selected lepton pairs compared to a combination of \PYTHIA6 and \textsc{hijing} (\PYTHIA6+\textsc{hijing}) Monte Carlo (MC) simulations. The $\Pp\mathrm{N}\to\cPZ\to\ell\ell$ process is simulated using the \PYTHIA6~\cite{Sjostrand:2006za} generator (version~6.424, tune~Z2~\cite{Field:2010bc}) with a mixture of pp and pn interactions corresponding to pPb collisions. Each \PYTHIA6 signal event is embedded in a minimum bias pPb background event which is produced with the \textsc{hijing} event generator version 1.383~\cite{Gyulassy:1994ew}. The detector response for each produced event is simulated with \GEANTfour~\cite{Agostinelli:2002hh}. The signal and background events have the same generated vertex location and are boosted to have the correct rapidity distribution in the laboratory frame. The embedding is done at the level of detector hits and then the events are processed through the trigger emulation and the event reconstruction chains. The reconstructed longitudinal primary vertex and overall multiplicity distributions are reweighted to match those observed in data.

An electron energy scale correction is extracted by fitting the energy to momentum ratio of electrons in a very pure $\PW\to\Pe\nu$ control sample~\cite{Chatrchyan:2013dga}. After fixing the shape of the distribution from MC, the energy to momentum ratio in data is fitted to derive the difference of the energy scale between data and MC, and then the data is corrected for this difference. A correction of the electron energy resolution is applied to MC by comparing the mass distribution of electron pairs between data and MC. Such corrections are also estimated for the $\cPZ\to\Pgmp\Pgmm$  channel and found to be negligible.

The raw yield, $S$, of Z boson candidates in the pPb sample is determined by counting the number of oppositely-charged lepton pairs in the 60--120\GeVcc mass region that fulfill the acceptance and quality requirements. This number is found to be 2183 in the muon channel and 1571 in the electron channel. The difference between the two channels is due to the tighter selection criteria applied to the electrons in order to suppress the higher background. A charge misidentification correction of 1\% is applied to the dielectron yields; this correction is negligible for dimuons. No events are found with more than one Z~boson candidate. For the differential cross sections, the measurement is performed in the dilepton transverse momentum or rapidity bins, where the rapidity is calculated in the center-of-mass frame.

Possible background contributions to the $\cPZ\to\ell\ell$ production are QCD multijet events, \ttbar pairs and electroweak processes such as W+jets, diboson (WW, WZ, ZZ), and $\cPZ\to\tau\tau$ production. Although the expected background contamination is small, an estimate based on data is used to subtract its contribution from the dilepton raw yield. For \ttbar, \bbbar, WW, and $\cPZ\to\tau\tau$ processes, two electron-muon events are expected for each dimuon or dielectron event, because of lepton universality. In the Z~boson mass range, the oppositely-charged electron-muon pairs are counted and translated into the expected number of muon or electron pairs, taking into account the differences in the muon and electron reconstruction and selection efficiencies. This background is subtracted from the dilepton raw yield and accounts for the main electroweak and \ttbar backgrounds, as well as for the part of QCD multijet background (such as \bbbar decays) that produces oppositely-charged leptons. The background from random combinations of other leptons in the event is estimated by counting the same-charge pairs. Additional electroweak contributions from W+jets and diboson production are found to be negligible via MC simulations. The fraction of background events subtracted from the raw yield is 2.4\% (2.9\%) in the muon (electron) channel, where the dominant background contribution comes from QCD processes, since no isolation requirements are imposed on the leptons.

\subsection{Efficiency and acceptance}

The efficiency, $\epsilon$, for Z bosons is defined as the number of reconstructed Z candidates, where both leptons fulfill the acceptance and quality requirements, divided by the number of generated Z bosons where both leptons fulfill the acceptance requirements. This combined reconstruction, lepton identification, and trigger efficiency is calculated from the \PYTHIA6\textsc{+hijing} simulation samples so that the effects of the pPb environment are taken into account.

For the rapidly falling dilepton \pt spectrum, an unfolding technique based on the inversion of a response matrix similar to the one used in Ref.~\cite{Chatrchyan:2011wt} is first applied to the data before applying the efficiency correction. The response matrix is constructed from the \PYTHIA6\textsc{+hijing} simulation to take into account the detector resolution effects.
The dilepton \pt resolution is about 0.5--1.5\GeVc, which results in a maximum bin-to-bin spill of about 30\% in the lowest \pt bins chosen for this analysis.
In the measurement of the dilepton rapidity, the unfolding is not necessary as the shape of the $y_\mathrm{cm}$ spectrum is almost flat and the resolution is a small fraction of the analysis bin size. Instead, the resolution effects in rapidity are taken into account in the efficiency corrections.

In order to correct for possible differences between data and simulation, a method derived from data is used to determine correction factors to the baseline efficiency from simulation. These correction factors are determined as a function of lepton $\eta$ and \pt by applying the \emph{tag-and-probe} method to both data and simulation to calculate single lepton efficiencies for reconstruction, identification, and triggering, similar to the method described in Ref.~\cite{Chatrchyan:2012xi}. The ratio of each efficiency from data over the corresponding efficiency in the simulation is then applied to reweight the simulation on a lepton-by-lepton basis. The efficiency for the Z~bosons, after correcting for the small differences between data and simulation, is found to be $0.878\pm0.015$ in the dimuon and $0.605\pm0.015$ in the dielectron decay channel. The sources of systematic uncertainties are described in Section~\ref{sec:syst}.

The acceptance, $\alpha$, is defined as the number of generated dilepton events where both leptons fulfill the acceptance requirements ($\pt^{\ell} > 20$\GeVc, $\abs{\eta_{\text{lab}}^{\ell}} < 2.4$) divided by the number of all generated dilepton events in the 60--120\GeVcc mass range. It is calculated using simulated events. The event generation is provided by the \textsc{powheg} generator~\cite{Nason:2004rx, Frixione:2007vw, Alioli:2008gx, Alioli:2010xd} with the CT10 free proton PDF set~\cite{Lai:2010vv}, interfaced with \PYTHIA6 parton shower, and the events are boosted to the laboratory frame (\textsc{powheg+}\PYTHIA6). Final-state photon radiation is also simulated by \PYTHIA6. The integrated acceptance is found to be $0.516\pm0.026$ in both decay channels.

\subsection{Systematic uncertainties}
\label{sec:syst}

The total systematic uncertainty in the Z boson production cross section is calculated by adding in quadrature the different contributions from the background subtraction, acceptance and efficiency determination, and the unfolding technique. The integrated luminosity, calibrated by the van der Meer scans~\cite{CMS:2014uoa}, has a systematic uncertainty of 3.5\%. It is the dominant systematic uncertainty of the measurement in the fiducial region.

The signal yield of Z candidates is affected by the uncertainty in the background subtraction method. The number of subtracted background events determined by the electron-muon method is varied conservatively by $\pm$100\% to assign an uncertainty in the signal yield. The uncertainty in the signal yield from this background variation is 1.7\% (1.8\%) in the muon (electron) channel.

The uncertainty in the correction factor for the electron energy scale is propagated as a systematic uncertainty in the dielectron yield. It is estimated to be 0.5\% in the inclusive yield and varies across the analysis \pt bins between 4 and 19\%. The residual difference in the mass resolution between data and simulation is taken as the systematic uncertainty in the electron channel. After propagating to the inclusive cross section, it accounts for a 1.1\% uncertainty.

The systematic uncertainty in the efficiency comes from two different sources. The first one is the uncertainty in the underlying rapidity and transverse momentum distributions reflecting the poorly known PDFs. This is estimated by applying a weight to the generated events that varies linearly between 0.7 and 1.3 over the $-3<y_\mathrm{cm}<3$ range, and a weight that varies between 0.9 and 1.1 over the $0<\pt<150$\GeVc range. These variations cover the predicted nuclear effects to the rapidity and \pt spectrum from different groups~\cite{Paukkunen:2010qg, Kang:2012am, Guzey:2012jp} as well as the statistical uncertainties in the present measurement and result in a 0.2\% uncertainty in the dilepton efficiency. Second, the statistical uncertainty in the correction factors coming from the ratio of data and simulation in the tag-and-probe method is propagated to the dilepton efficiency. In addition, the tag-and-probe technique itself carries an uncertainty of about 1\%, estimated from differences observed in the efficiencies by varying the functional form or the range of the fits. Finally, the uncertainties in the three different components of the efficiency are combined in quadrature, resulting in an overall uncertainty in the dimuon (dielectron) efficiency of 1.7\% (2.5\%).

All the uncertainties above are evaluated in bins of dilepton rapidity and transverse momentum to give uncertainties in the differential cross sections. The systematic uncertainty of the forward-backward asymmetry is calculated from the rapidity differential cross section. The uncertainties in the background, electron energy scale, and efficiency are propagated without assuming any cancellation. The uncertainty of the luminosity cancels in the ratio.

There is an additional uncertainty in the dilepton \pt spectrum coming from the matrix inversion procedure used for the unfolding. This uncertainty is determined by varying the generated dilepton \pt distribution and the single lepton \pt resolution. The reconstructed \pt distributions from \PYTHIA6\textsc{+hijing} and \textsc{powheg+}\PYTHIA6, as well as the weighted \pt spectrum reflecting possible nPDF differences, are all studied and their effect on the results is directly evaluated. These two sources give a combined uncertainty in the unfolded yield of about 1--5\%, depending on the \pt bin.

The uncertainty due to the acceptance correction is estimated by changing the shape of the generated rapidity and \pt distributions of the Z bosons with the same functions as described for the efficiency uncertainty in order to cover differences in PDFs and possible nuclear effects. The resulting uncertainty in the acceptance is about 5\% from the extrapolation to the most forward and backward rapidity regions and it only affects the total cross section. Table~\ref{tab:syst} summarizes the systematic uncertainties in the two decay channels.

\begin{table}[thb]
\centering
\topcaption{Summary of systematic uncertainties in the two decay channels.}
\label{tab:syst}
\begin{tabular}{l c c}
\hline
Source & $\cPZ\to\mu\mu$ & $\cPZ\to\Pe\Pe$ \\
\hline
Background & 1.7\% & 1.8\% \\
Electron energy scale & --- & 0.5\% \\
Electron resolution & --- & 1.1\% \\
Efficiency & 1.7\% & 2.5\% \\
Unfolding of \pt spectrum & \multicolumn{2}{c}{1--5\%} \\
\hline
Acceptance & \multicolumn{2}{c}{5\%} \\
\hline
Luminosity & \multicolumn{2}{c}{3.5\%} \\
\hline
Total (fiducial cross section) & 4.2\% & 4.8\% \\
Total (total cross section) & 6.6\% & 6.9\% \\
\hline
\end{tabular}
\end{table}

\section{Results}

The results are primarily compared to the NLO pp predictions from the \textsc{powheg+}\PYTHIA~6 generator using the CT10~\cite{Lai:2010vv} free proton PDF set.
The $\Pp\mathrm{N}\to\cPZ\to\ell\ell$ process is also simulated with the \textsc{mcfm}~\cite{Campbell:2011bn} generator (version 6.7) using the CT10 free proton PDF set, as well as the EPS09~\cite{Eskola:2009uj} and DSSZ~\cite{deFlorian:2011fp} nuclear PDF sets.
Since these predictions include theoretical uncertainties, their statistical compatibility with the measurements can be tested.
All predictions are scaled by the number of nucleons in the Pb nucleus ($A=208$) as is expected in the case of elementary nucleon-nucleon collision scaling.

The cross section of Z boson production is calculated using Eq.~(\ref{eq:xsec}) for both decay channels. The analysis of the muon channel results in a fiducial cross section ($\pt^{\ell}>20\GeVc,\abs{\eta_{\text{lab}}^{\ell}}<2.4$) of $70.1 \pm 1.5\stat\pm 1.7\syst\pm 2.5\lum\unit{nb}$ and the electron channel gives $73.9 \pm 1.9\stat\pm 2.8\syst\pm 2.6\lum\unit{nb}$.

The muon and electron results, which agree within statistical and systematic uncertainties, are combined, separating out the uncertainty related to the integrated luminosity. The best linear unbiased estimate (BLUE) technique~\cite{Lyons:1988rp} is applied, taking the muon and electron channel cross sections and their uncertainties in each bin to be uncorrelated.

The measured inclusive Z boson production cross section in the fiducial region, where both leptons fulfill the acceptance requirements is
\begin{linenomath}
\ifthenelse{\boolean{cms@external}}{
\begin{multline}
\sigma_{\mathrm{pPb}\to\cPZ\to\ell\ell}(\pt^{\ell}>20\GeVc, \abs{\eta_{\text{lab}}^{\ell}}<2.4) \\= 71.3 \pm 1.2\stat\pm 1.5\syst\pm 2.5\lum\unit{nb}.
\end{multline}
}{
\begin{equation}
\sigma_{\mathrm{pPb}\to\cPZ\to\ell\ell}(\pt^{\ell}>20\GeVc, \abs{\eta_{\text{lab}}^{\ell}}<2.4) = 71.3 \pm 1.2\stat\pm 1.5\syst\pm 2.5\lum\unit{nb}.
\end{equation}
}
\end{linenomath}
The \textsc{powheg+}\PYTHIA6 prediction gives a Z~boson cross section in pp collisions at $\sqrt{s}=5.02$\TeV of $338\pm17$\unit{pb} for $\cPZ\to\ell\ell$ production in the 60--120\GeVcc mass range after applying the acceptance requirements on the leptons. The uncertainties in the theoretical prediction in pp collisions amount to about 5\% and arise from missing higher-order corrections and from the uncertainties in the PDF sets. Scaling the pp cross section by $A=208$, results in the prediction of $70.4\pm3.5$\unit{nb} for the pPb cross section, which is consistent with the measured value.

For the acceptance-corrected total cross section, the systematic uncertainty in the acceptance is correlated between the two decay channels, which is taken into account in the BLUE method. The combined total Z boson production cross section in the 60--120\GeVcc mass region is
\begin{linenomath}
\begin{equation}
\sigma_{\mathrm{pPb}\to\cPZ\to\ell\ell} = 138.1 \pm 2.4\stat\pm 8.6\syst\pm 4.8\lum\unit{nb}.
\end{equation}
\end{linenomath}
This measurement has an uncertainty of about 5\% from the extrapolation of the detector acceptance to the full phase space. The \textsc{powheg+}\PYTHIA6 generator after scaling predicts $136.1\pm6.8$\unit{nb}, which is consistent with the measured value.

\begin{figure}[thb]
 \centering
    \includegraphics[width=0.46\textwidth]{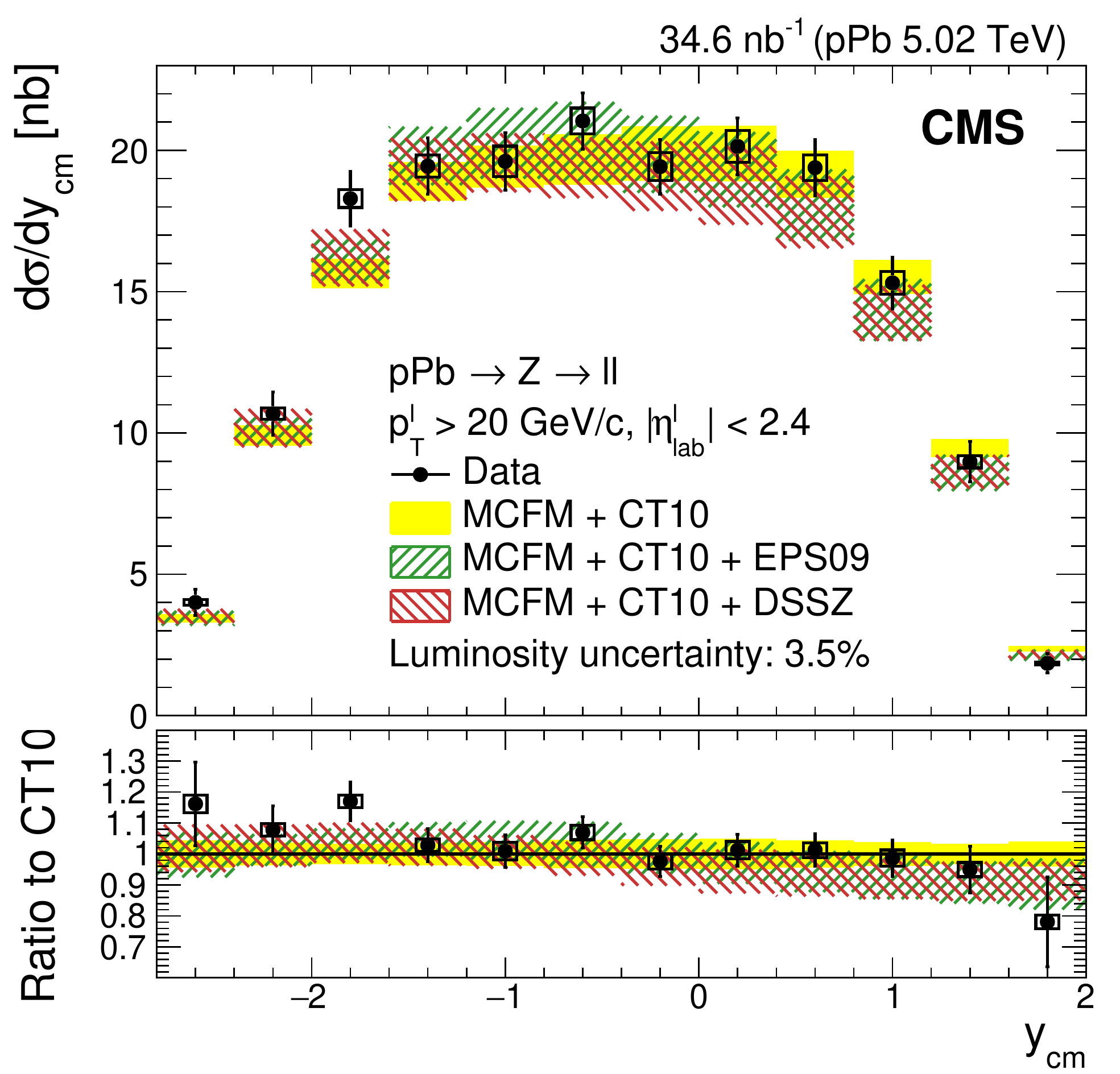}
    \caption{Differential cross section of the Z bosons in pPb collisions as a function of rapidity in the fiducial region for the combined leptonic decay channel. Colored boxes are predictions from the \textsc{mcfm} generator, scaled by 208 (see text), and using nuclear (EPS09 and DSSZ) or free (CT10) PDF sets. The bottom panel shows the ratio of the data and the nPDF predictions to the CT10 PDF set. The vertical bars (boxes) represent the statistical (systematic) uncertainties.}
    \label{fig:sigmay}
\end{figure}

Figure~\ref{fig:sigmay} shows the differential cross section of the Z bosons in the fiducial region in pPb collisions as a function of rapidity.
The luminosity normalization uncertainty of 3.5\% is not shown.
The \MCFM theoretical predictions, both with and without nuclear modification, are consistent with the measured differential cross section within uncertainties. The corresponding rapidity dependence predicted by \POWHEG{}+\PYTHIA6 for pp collisions agrees with the \MCFM calculation for pN collisions using the CT10 PDF set without nuclear modification, showing that any dependences on isospin or the PDF set are within the theoretical uncertainties.

\begin{figure}[thb]
 \centering
    \includegraphics[width=0.46\textwidth]{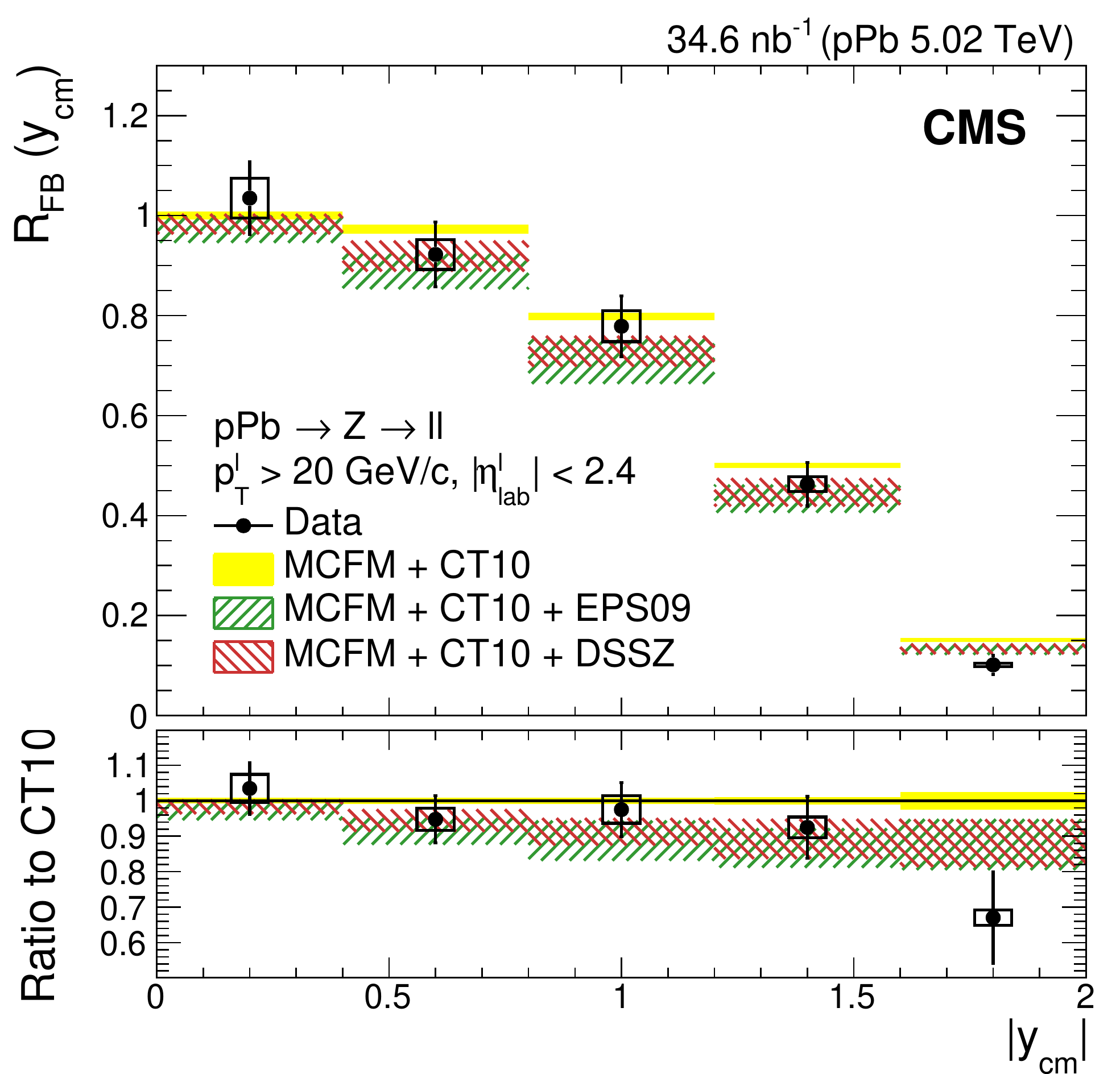}
    \caption{Forward-backward asymmetry $R_\mathrm{FB}$ distribution of the Z bosons in pPb collisions as a function of rapidity in the fiducial region for the combined leptonic decay channel compared to the predictions from the \textsc{mcfm} generator with nuclear (EPS09 and DSSZ) or free (CT10) PDF sets. The bottom panel shows the ratio of the data and the nPDF predictions to the CT10 PDF set. The vertical bars (boxes) represent the statistical (systematic) uncertainties.}
    \label{fig:fwdbwd} 
\end{figure}

Nuclear effects are expected to modify the rapidity distribution asymmetrically and thus they can be further quantified by the forward-backward asymmetry defined in Eq.~(\ref{eq:fb}). This quantity is expected to be more sensitive to nuclear effects~\cite{Paukkunen:2010qg} because normalization uncertainties cancel both in theory and in experiment. Figure~\ref{fig:fwdbwd} shows the measured forward-backward asymmetry as a function of $\abs{y_\mathrm{cm}}$ compared to the \textsc{mcfm} predictions with and without nuclear modification.

While being consistent with the three theoretical predictions shown, the data tend to favor the presence of nuclear effects in PDFs. The ATLAS collaboration reached similar conclusions from their Z boson measurement~\cite{Aad:2015gta}. Together with the measured W~boson production in pPb collisions~\cite{Khachatryan:2015hha}, these results can reduce the nPDF uncertainties by adding new data to the global fits in a previously unexplored region of the $(Q^2$, $x)$ phase space.

In order to quantify the agreement between the measurements and the predictions with the different PDF sets, a $\chi^2$ test is performed for the rapidity-dependent differential cross section and the forward-backward asymmetry. The few correlations in the experimental uncertainties, only relevant for the cross section but not for the asymmetry, are taken into account, as well as the correlations in the theoretical uncertainties. The resulting $\chi^2$ values and probabilities are given in Table~\ref{tab:chi2}. The theoretical calculations including nuclear effects provide a somewhat better description of the measurements.

\begin{table*}[thb]
\centering
\topcaption{Results of the $\chi^2$ test between the measurements and the theoretical predictions with and without nuclear modification from the EPS09 or DSSZ nPDF sets. The differential cross section and the forward-backward asymmetry have twelve and five numbers of degrees of freedom (NDF), respectively. \label{tab:chi2}}
\renewcommand{\arraystretch}{1.3}
\begin{tabular}{ccccccc}
\hline
\multirow{2}{*}{Observable} & \multicolumn{2}{c}{CT10} & \multicolumn{2}{c}{CT10+EPS09} & \multicolumn{2}{c}{CT10+DSSZ} \\
\cline{2-7}
 & $\chi^2/$NDF & Probability & $\chi^2/$NDF & Probability & $\chi^2/$NDF & Probability \\
\hline
$\rd\sigma/\rd y_\mathrm{cm}$ & 10.8/12 & 54\% & 7.4/12 & 83\% & 6.6/12 & 88\% \\
$R_\mathrm{FB}$ & 7.3/5  & 20\% & 3.9/5 & 56\% & 3.4/5 & 64\% \\
\hline
\end{tabular}
\end{table*}

\begin{figure}[t]
 \centering
    \includegraphics[width=0.46\textwidth]{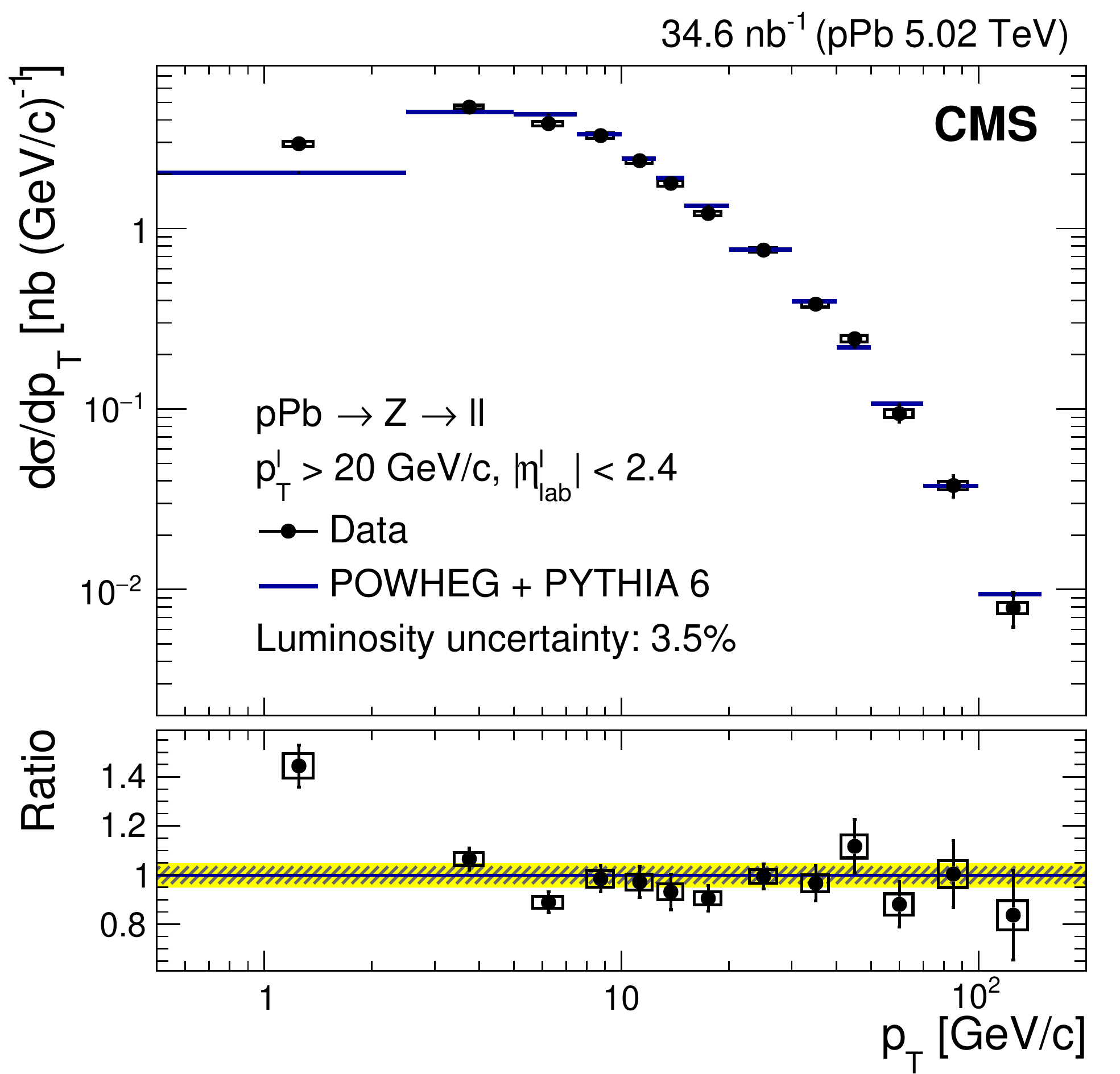}
    \caption{Differential cross section of the Z bosons in pPb collisions as a function of transverse momentum in the fiducial region for the combined leptonic decay channel compared to the prediction from the \textsc{powheg+}\PYTHIA6 generator scaled by the number of nucleons in the Pb nucleus. The vertical bars (boxes) represent the statistical (systematic) uncertainties. The 3.5\% luminosity uncertainty is shown in the ratio plot as a hashed band together with the assumed 5\% theoretical uncertainty, shown as a yellow band.}
    \label{fig:sigmapt}
\end{figure}

Figure~\ref{fig:sigmapt} shows the differential cross section as a function of $\pt$ in the fiducial region. The results are compared only to theoretical predictions from \textsc{powheg+}\PYTHIA6, because the expected nuclear modification of the \pt spectrum is small compared to the uncertainties in the theory~\cite{Kang:2012am, Guzey:2012jp}. No large deviations are found from the theoretical cross sections, apart from the lowest dilepton \pt bins where the differences from \textsc{powheg+}\PYTHIA6 are similar to the ones observed in the pp measurements at 7\TeV~\cite{Chatrchyan:2011wt, Aad:2011gj}.

\section{Summary}

The cross section of Z boson production has been measured in the muon and electron decay channels in pPb collisions at $\sqrt{s_\mathrm{NN}} = 5.02$\TeV. The NLO pp inclusive cross section from \textsc{powheg+}\PYTHIA6 scaled by the number of elementary nucleon-nucleon collisions is in agreement with the measured pPb cross section. The pPb theoretical predictions for the differential cross section as a function of the Z boson rapidity with and without nuclear effects are compared to the measurement. Given the small differences in these predictions and their inherent theoretical uncertainties as well as the sensitivity of the data, both scenarios, presence or not of nuclear effects, are consistent with the data. A more sensitive variable, the forward-backward asymmetry, deviates from predictions assuming free proton PDFs by an amount which is compatible with both the EPS09 and the DSSZ nPDF modifications, although the statistical precision of the measurement precludes making a definitive statement. The differential cross section as a function of the Z boson transverse momentum has been measured and is found to be in agreement with pp predictions from \textsc{powheg+}\PYTHIA6, except at very low transverse momentum, where similar deviations as previously seen in pp are observed. The presented results provide new data for constraining nuclear PDF fits.

\begin{acknowledgments}
We congratulate our colleagues in the CERN accelerator departments for the excellent performance of the LHC and thank the technical and administrative staffs at CERN and at other CMS institutes for their contributions to the success of the CMS effort. In addition, we gratefully acknowledge the computing centers and personnel of the Worldwide LHC Computing Grid for delivering so effectively the computing infrastructure essential to our analyses. Finally, we acknowledge the enduring support for the construction and operation of the LHC and the CMS detector provided by the following funding agencies: BMWFW and FWF (Austria); FNRS and FWO (Belgium); CNPq, CAPES, FAPERJ, and FAPESP (Brazil); MES (Bulgaria); CERN; CAS, MoST, and NSFC (China); COLCIENCIAS (Colombia); MSES and CSF (Croatia); RPF (Cyprus); MoER, ERC IUT and ERDF (Estonia); Academy of Finland, MEC, and HIP (Finland); CEA and CNRS/IN2P3 (France); BMBF, DFG, and HGF (Germany); GSRT (Greece); OTKA and NIH (Hungary); DAE and DST (India); IPM (Iran); SFI (Ireland); INFN (Italy); MSIP and NRF (Republic of Korea); LAS (Lithuania); MOE and UM (Malaysia); CINVESTAV, CONACYT, SEP, and UASLP-FAI (Mexico); MBIE (New Zealand); PAEC (Pakistan); MSHE and NSC (Poland); FCT (Portugal); JINR (Dubna); MON, RosAtom, RAS and RFBR (Russia); MESTD (Serbia); SEIDI and CPAN (Spain); Swiss Funding Agencies (Switzerland); MST (Taipei); ThEPCenter, IPST, STAR and NSTDA (Thailand); TUBITAK and TAEK (Turkey); NASU and SFFR (Ukraine); STFC (United Kingdom); DOE and NSF (USA).

Individuals have received support from the Marie-Curie program and the European Research Council and EPLANET (European Union); the Leventis Foundation; the A. P. Sloan Foundation; the Alexander von Humboldt Foundation; the Belgian Federal Science Policy Office; the Fonds pour la Formation \`a la Recherche dans l'Industrie et dans l'Agriculture (FRIA-Belgium); the Agentschap voor Innovatie door Wetenschap en Technologie (IWT-Belgium); the Ministry of Education, Youth and Sports (MEYS) of the Czech Republic; the Council of Science and Industrial Research, India; the HOMING PLUS program of the Foundation for Polish Science, cofinanced from European Union, Regional Development Fund; the OPUS program of the National Science Center (Poland); the Compagnia di San Paolo (Torino); MIUR project 20108T4XTM (Italy); the Thalis and Aristeia programs cofinanced by EU-ESF and the Greek NSRF; the National Priorities Research Program by Qatar National Research Fund; the Rachadapisek Sompot Fund for Postdoctoral Fellowship, Chulalongkorn University (Thailand); the Chulalongkorn Academic into Its 2nd Century Project Advancement Project (Thailand); and the Welch Foundation, contract C-1845.
\end{acknowledgments}
\bibliography{auto_generated}

\cleardoublepage \appendix\section{The CMS Collaboration \label{app:collab}}\begin{sloppypar}\hyphenpenalty=5000\widowpenalty=500\clubpenalty=5000\textbf{Yerevan Physics Institute,  Yerevan,  Armenia}\\*[0pt]
V.~Khachatryan, A.M.~Sirunyan, A.~Tumasyan
\vskip\cmsinstskip
\textbf{Institut f\"{u}r Hochenergiephysik der OeAW,  Wien,  Austria}\\*[0pt]
W.~Adam, E.~Asilar, T.~Bergauer, J.~Brandstetter, E.~Brondolin, M.~Dragicevic, J.~Er\"{o}, M.~Flechl, M.~Friedl, R.~Fr\"{u}hwirth\cmsAuthorMark{1}, V.M.~Ghete, C.~Hartl, N.~H\"{o}rmann, J.~Hrubec, M.~Jeitler\cmsAuthorMark{1}, V.~Kn\"{u}nz, A.~K\"{o}nig, M.~Krammer\cmsAuthorMark{1}, I.~Kr\"{a}tschmer, D.~Liko, T.~Matsushita, I.~Mikulec, D.~Rabady\cmsAuthorMark{2}, N.~Rad, B.~Rahbaran, H.~Rohringer, J.~Schieck\cmsAuthorMark{1}, R.~Sch\"{o}fbeck, J.~Strauss, W.~Treberer-Treberspurg, W.~Waltenberger, C.-E.~Wulz\cmsAuthorMark{1}
% ...
\vskip\cmsinstskip
% ...
1:~~Also at Vienna University of Technology, Vienna, Austria\\
2:~~Also at CERN, European Organization for Nuclear Research, Geneva, Switzerland\\
% ...

\end{sloppypar}
\end{document}